\documentclass[a4paper]{article}

\usepackage{INTERSPEECH2020}
\usepackage{cleveref}
\usepackage{graphicx}
\usepackage{booktabs}

\title{Speech Recognition and Multi-Speaker Diarization of Long Conversations}
\name{Huanru Henry Mao, Shuyang Li, Julian McAuley, Garrison W. Cottrell}
\address{UC San Diego}
\email{\{hhmao, shl008, jmcauley, gary\}@eng.ucsd.edu}

\begin{document}

\maketitle
\begin{abstract}
Speech recognition (ASR) and speaker diarization (SD) models have traditionally been trained separately to produce rich conversation transcripts with speaker labels.
Recent advances \cite{DBLP:conf/interspeech/ShafeySS19} have shown that joint ASR and SD models can learn to leverage audio-lexical inter-dependencies to improve word diarization performance. 
We introduce a new benchmark of hour-long podcasts collected from the weekly \textit{This American Life} radio program to better compare these approaches when applied to extended multi-speaker conversations.
We find that training separate ASR and SD models perform better when utterance boundaries are known but otherwise joint models can perform better.
To handle long conversations with unknown utterance boundaries, we introduce a striding attention decoding algorithm and data augmentation techniques which, combined with model pre-training, improves ASR and SD.
\end{abstract}

\section{Introduction}
Automatic speech recognition (ASR) and speaker diarization (SD) of natural conversation are tasks of broad interest with applications including transcribing meetings, phone calls, and interviews, among others.
Traditionally, ASR and SD systems each operate independently on acoustic information to generate transcript text and label speaker segments.
These outputs are then reconciled into a final speaker-annotated transcript.
A limitation of training independent ASR and SD systems is that the models are unable to leverage the inter-dependencies between these two predictive tasks.
For example, lexical cues from transcripts can help improve speaker turn change prediction \cite{DBLP:conf/interspeech/ParkG18}.

Recent work has shown promising results in learning sequence transduction models that jointly perform ASR and SD in a two-speaker clinical setting by simply adding a speaker change token to the model's vocabulary \cite{DBLP:conf/interspeech/ShafeySS19}.
To further explore these types of end-to-end approaches, we expand the joint framework to encompass ASR and SD in an open-domain setting for extended multi-speaker conversations.
We introduce a benchmark dataset for this setting, consisting of 663 podcast episodes and transcripts collected from the weekly \textit{This American Life} (TAL) radio program.\footnote{Data and code can be found on: https://github.com/calclavia/tal-asrd}
TAL is unique in two ways: each episode is an hour-long conversation and contains an average of 18 unique speakers in three roles.
We propose two tasks for joint ASR and diarization: TAL aligned and unaligned, to evaluate models under situations where utterance bounds are either provided or unknown respectively.
To benchmark performance in each setting, we measure the transcription error via word error rate (WER) and introduce a new metric, multi-speaker word diarization error (MWDE), to evaluate word-level speaker alignment.
MWDE generalizes the previously proposed two-speaker word diarization error rate \cite{DBLP:conf/interspeech/ShafeySS19} to multiple speakers.

We compare training separate ASR and SD models against the joint framework \cite{DBLP:conf/interspeech/ShafeySS19} for our multi-speaker setting and find that the separate framework is superior when known utterance boundaries are provided but worse otherwise, suggesting that joint models may be more appropriate as fully end-to-end rich transcription systems.
To handle long conversations, we introduce a striding attention decoding algorithm  to adapt a model trained on TAL aligned to the hour-long unaligned setting.
We propose pre-training and data augmentation methods to complement the algorithm, achieving 16.1\% and 15.8\% absolute improvements to WER and MWDE on TAL unaligned.

\begin{figure}[t]
  \centering
  \includegraphics[width=1\linewidth]{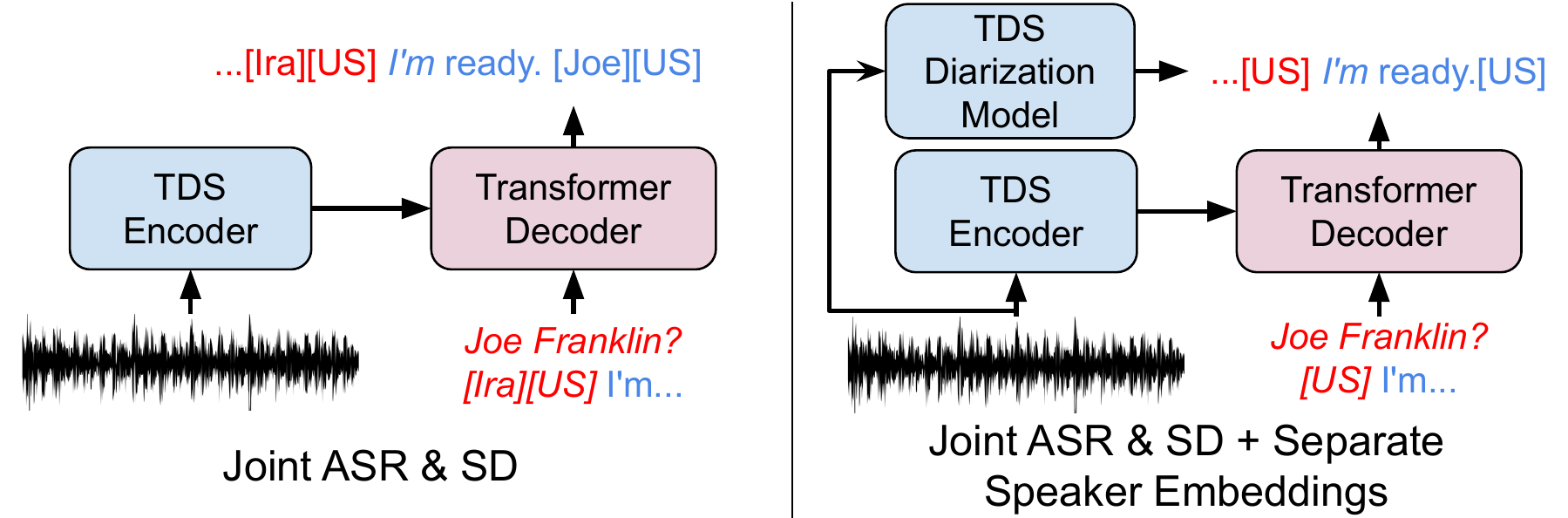}
  \caption{
  The joint model concatenates utterances into a sequence delimited by a speaker token (e.g.,~[Ira]) and an utterance separator [US] token.
  Alternatively, separate speaker embeddings can be used instead of speaker tokens for diarization.
  }
  \label{fig:model}
  \vspace{-0.2in}
\end{figure}

\section{Dataset}


We collected podcast episodes from the \textit{This American Life} radio program from 1995 to 2020, comprising 701 episodes which we cleaned and processed for a total of 663 episodes (38 had alignment issues due to inserted ads) and 637.70 hours of audio.
Each episode corresponds to a ``single conversation", and TAL comprises hundreds of lengthy dialogs.
On average, each conversation is 58 minutes long, consisting of 247 dialog turns between 18 different speakers.
We additionally collected professionally transcribed, publicly-available transcripts for each episode, which are aligned at the utterance-level.\footnote{e.g.,~https://www.thisamericanlife.org/74/transcript}
Each utterance comprises an average of 45 words and 3 sentences, for a total of 7,390,793 words and 520,676 sentences across 163,808 utterances.
90\% of utterances fall just under 30 seconds in duration and 100 words in length.


Conversations are loosely organized around a theme (e.g.,~``Middle School") with several guests who tell stories and engage with the host.
There are 6,608 unique speakers identified over all episodes, with each episode featuring an average of 18 unique interlocutors, posing a challenge for SD systems.
These speakers have been annotated with role labels, with each speaker acting as a ``host", ``interviewer", or ``subject".
Hosts tend toward expository speech with long turns of dialog and are responsible for 42.4\% of utterances.
Interviewers pose questions to facilitate discussion, speaking 12.8\% of the time with shorter utterances.
Subjects make up the remaining 44.8\% of utterances and generally speak at length.
The subject matter and dialog structure also poses a challenge for transcription, with a vocabulary of 1,309,647 unique words and references to 53,792 unique named entities as identified via \texttt{spaCy}.\footnote{https://spacy.io/, \texttt{en\_core\_web\_sm} model}

We compare TAL to several benchmark datasets for ASR and SD in \Cref{tab:dataset-comparison}, alongside the clinical dataset used in \cite{DBLP:conf/interspeech/ShafeySS19}.
A large body of research in ASR has focused on the 1,000-hour LibriSpeech dataset \cite{DBLP:conf/icassp/PanayotovCPK15} of audiobook segments, while SD research focuses on telephone conversation transcripts from the Fisher \cite{cieri2004fisher}, CALLHOME \cite{canavan1997callhome} and Switchboard \cite{godfrey1992switchboard} corpora.
Of these datasets, only LibriSpeech and TAL are free and openly accessible.
We note that RadioTalk \cite{DBLP:conf/interspeech/BeefermanBR19} also collected a large corpus of radio program transcripts, but did so using a noisy automated system with no corresponding gold labels and did not release the audio.
In contrast to other ASR datasets, TAL transcripts contain proper punctuation and casing.
Professional transcribers for TAL may elect to ignore stutters and irrelevant repetitions, performing minor grammatical fixes to the spoken words.
Thus, transcription models for this setting must capture higher-level semantics of the utterance.
TAL also contains a diverse set of speaker accents, varying rates of speech, and background music, making it an acoustically challenging dataset.

We standardized TAL's raw audio by preprocessing all audio to 16KHz mono-channel wav format.
Approximately 9\% of the publicly released transcripts for TAL contain alignment errors, primarily stemming from advertising preceding each act.
We manually checked the episodes, discarding 38 episodes with content-related errors and manually re-aligned and re-transcribed another 25 episodes.
We split these cleaned conversations into disjoint training, validation, and test sets comprising 593, 34, and 36 episodes respectively.

\begin{table}[t]
\centering
\caption{Comparison of benchmark datasets for ASR and SD. MPC: minutes/conversation; SPC: speakers/conversation.}
\label{tab:dataset-comparison}
\begin{tabular}{@{}lrrrrl@{}}
\toprule
 &
  \multicolumn{1}{l}{\textbf{Conv}} &
  \multicolumn{1}{l}{\textbf{Hours}} &
  \multicolumn{1}{l}{\textbf{MPC}} &
  \multicolumn{1}{l}{\textbf{SPC}} &
  \textbf{Setting} \\ \midrule
LibriSpeech &
  \multicolumn{1}{c}{--} &
  1K &
  \multicolumn{1}{c}{--} &
  1 &
  Book \\
CALLHOME    & 120   & 60   & 30.0 & 2  & Phone   \\
Switchboard & 2.4K  & 260  & 6.5  & 2  & Phone   \\
Fisher      & 16.4K & 2.7K & 6.0  & 2  & Phone   \\
Clinical \cite{DBLP:conf/interspeech/ShafeySS19}      & 100K  & 15K  & 6.7  & 2  & Medical \\ \midrule
TAL (Ours)  & 663   & 637  & \textbf{57.7} & \textbf{18} & Podcast \\ \bottomrule
\end{tabular}
\vspace{-0.15in}
\end{table}

\section{Task}
We present the TAL aligned and unaligned tasks to test a model's capability to diarize and transcribe text under bounded and unbounded conditions.
The \textbf{TAL Aligned Task} measures ASR and SD performance when utterance bounds are provided.
Given a single utterance of input speech $X=(x_1,\ldots,x_n)$ (in raw waveform or spectrogram format), a model must produce a sequence of vocabulary tokens $Y = (y_1,\ldots,y_m)$ and speaker labels $S = (s_1, \ldots, s_m)$ where $y_i \in V$ (vocabulary) and $s_i \in H$ (speaker IDs). 
We set the beginning and terminal tokens $y_1, y_m$ as the special utterance separator \texttt{[US]} token.
We only consider utterances between 3 to 30 seconds, comprising 6,774 utterances in the test set.
The \textbf{TAL Unaligned Task} is similar to the above, but utterance bounds are not provided, forcing the model to conduct full-conversation ASR and SD.
An hour-long podcast episode is provided as $X$, and the target outputs as $Y$ and $S$, corresponding to the full episode transcript and gold speaker labels, respectively.
To perform well under this setting, the model must learn to determine utterance alignments.
This closely resembles real-world audio transcription without known segmentation.

ASR is evaluated using word error rate (WER), comparing the model's output tokens $\hat{Y}$ against reference tokens $Y$.
We retain casing and punctuation, calculating WER over model outputs tokenized via the Punkt tokenizer \cite{DBLP:journals/coling/KissS06}, with incorrectly cased words counted as errors.
All generated outputs that do not terminate (with the \texttt{[US]} token) are treated as 100\% WER.

Prior work in the joint ASR and SD setting evaluated diarization with WDER \cite{DBLP:conf/interspeech/ShafeySS19}, defined as $\frac{S_w + C_w}{S + C}$ where $S_w$ is the number of ASR substitutions with the wrong speaker label, $C_w$ is the number of correct transcriptions with the wrong speaker label, and $S$ and $C$ are the total number of substituted and correct words, respectively.
WDER was originally proposed for a doctor-patient setting where we care about absolute label correctness.
In diarization settings, however, the goal is speaker disambiguation, and different speakers \textit{within a single conversation} must have distinct labels (that need not match their ground truth identities).
Thus, while WDER can measure role classification error, it is an inappropriate metric for multi-speaker diarization error.
To measure the latter, we introduce a new metric: multi-speaker word diarization error (MWDE).
For MWDE we first compute the optimal alignment 
between output and reference speakers $m$ out of all possible alignments $M$---much like computing multi-speaker diarization error rate (DER) \cite{kuhn1955hungarian, DBLP:conf/interspeech/Galibert13a}---and then calculate WDER with the new alignments:
\begin{align}
    \text{MWDE} &= \min_{m \in M}{\text{WDER}_m}
\end{align}
Like WDER, MWDE does not account for ASR additions and deletions, due to ambiguous reference speaker labels.
Combined with WER, MWDE allows us to evaluate joint diarization and transcription systems based on word-level alignments, which is appropriate for practical applications.

\section{Approach}
We compare three different frameworks to perform ASR and diarization.
As it is computationally infeasible to directly train models on TAL unaligned, we train each of our models on TAL aligned and evaluate on both the aligned and unaligned test sets.

\subsection{Separate ASR and Diarization}
\label{sec:sep_framework}
Typically, ASR and SD on multi-speaker audio are conducted through independent pipelines.
The ASR model is trained to predict the spoken words $Y$ in the audio.
The SD model is trained to produce speaker embeddings which are then clustered during inference to determine who spoke when \cite{DBLP:conf/interspeech/YinBB18} and when the speakers change.
A reconciliation step is then required to assign the SD model's time-position speaker labels to the ASR model's output $Y$ to produce word-level speaker labels $S$.

For our baselines, we train a sequence transduction model to perform ASR.
We seek to learn hidden representations
$h_i = \text{dec}(\text{enc}(X), y_{< i})$
for each token output, where $\text{enc}(\ldots)$ and $\text{dec}(\ldots)$ refer to an encoder and decoder neural network respectively.
This representation is used to predict probabilities for each token
$P(y_i|X, y_{< i}) = \text{softmax}(W h_i)$
of the output vocabulary, where $W$ is a learned weight matrix.
We minimize the cross entropy loss of $P(y_i|X, y_{< i})$ against true tokens in a conventional sequence-to-sequence manner.
We train a diarization-only acoustic model to classify speakers \cite{DBLP:conf/interspeech/FujitaKHNW19} from the TAL training set.
We learn features for each audio frame, and calculate cross-entropy loss of predicted speaker against the true speaker for that frame.
During evaluation, we compute the weighted average features for each word using the \textit{attention focus} (Section \ref{sec:decoding}) and then cluster with HDBScan \cite{DBLP:conf/pakdd/CampelloMS13} to assign word-level speaker labels.

\subsection{Joint ASR and Diarization}
To explore an end-to-end approach to simultaneous ASR and SD, we can formulate our task as a joint sequence transduction task to jointly predict the spoken words, speaker identity and turn changes \cite{DBLP:conf/interspeech/ShafeySS19}.
Prior to each utterance termination token \texttt{[US]}, we insert the speaker ID token such that such that $Y_{\text{aug}}=(y_1,\ldots,h,y_m), y_{1\ldots m} \in V,h \in H$.
At test time, once the speaker token has been produced, all preceding tokens in the utterance are assigned to that speaker $s_i = h$.
To handle unseen speakers, SD systems typically use some form of clustering  \cite{DBLP:conf/icassp/Garcia-RomeroSS17, DBLP:conf/slt/SellG14,DBLP:conf/bmvc/ZhaiW19}.
Instead, we simply use the model's predicted speaker from the training set as our label for the unseen speaker.

\subsection{Boosting with Separate Diarization Model}
Recent work has shown the importance of lexical information in speaker change detection \cite{DBLP:conf/interspeech/ParkG18, DBLP:conf/cikm/MengMJ17}, but the same has not been shown for speaker identity prediction.
To investigate this, we explore an alternative approach where we use the above joint model to predict speaker change and a separate diarization model (from Section \ref{sec:sep_framework}) to produce speaker embeddings.
This setting differs from the separate framework in that the speaker change bounds are determined by the joint model and not the diarization model (\Cref{fig:model}).
We produce an utterance embedding by averaging all speaker embeddings that fall within an utterance, reducing the noise produced by individual embeddings.
We then use HDBScan \cite{DBLP:conf/pakdd/CampelloMS13} to cluster speakers.
We report diarization results via this ``boosting" with a separately-trained SD model in the \textbf{SD+} column of \Cref{tab:experiments}.

\subsection{Model Architecture}
While Transformer acoustic models have shown promising performance in ASR \cite{DBLP:journals/corr/abs-1910-09799}, in preliminary experiments we have found that they scale poorly to long sequences due to high memory requirements.
Instead, we use a Time-Depth Separable Convolution (TDS) \cite{DBLP:conf/interspeech/Hannun0XC19} acoustic model, which has a better computation to performance trade-off.
We followed the same configuration as TDS except the following.
Instead of an RNN, our decoder is a 4-layer Transformer \cite{DBLP:conf/nips/VaswaniSPUJGKP17} decoder with 512 hidden units per layer and 64-dimensional factorized token embeddings \cite{DBLP:journals/corr/abs-1909-11942}.
We replace all LayerNorm \cite{DBLP:journals/corr/BaKH16} with ReZero initialization \cite{DBLP:journals/corr/abs-2003-04887} and 2D convolutions with 1D convolutions \cite{pratap2020scaling}.
These modifications accelerated training, and our model achieves comparable ASR performance to the original \cite{DBLP:conf/interspeech/Hannun0XC19} on the LibriSpeech \textit{clean} (6.18\% vs. 5.58\%) and \textit{other} (15.62\% vs. 15.30\%) development sets.
We use the same architecture for all our experiments.
For training diarization-only models, we use only the TDS encoder without a decoder.

\subsection{Pre-training}
To boost our model's performance, we propose leveraging pre-training techniques, which have been shown to successfully improve natural language \cite{DBLP:conf/naacl/DevlinCLT19} and computer vision tasks \cite{DBLP:conf/cvpr/LongSD15}.
We pre-train the encoder \cite{DBLP:conf/slt/ToshniwalKCWSL18, DBLP:conf/Interspeech/SchneiderBCA19} via ASR on the LibriSpeech corpus, discarding the decoder module due to vocabulary mismatch and transferring the encoder's learned weights.

\subsection{Decoding Long Conversations}
\label{sec:decoding}
TAL aligned evaluation is straightforward, and we decode from our model following conventional approaches using beam search of size 5.
TAL unaligned evaluation requires decoding over an hour of audio.
We introduce an algorithm we call \textit{striding window attention} to enable our model to efficiently decode full hour-long episodes.
First, we run the full episode through the WebRTC\footnote{https://webrtc.org/} Voice Activity Detector (VAD) to remove non-speech segments and compute full-episode audio features using our TDS encoder.
Due to memory limitations 
we can only attend to a window of features covering a 30-second receptive field when generating a particular output.
The key challenge is to effectively stride this attention window to relevant audio segments without the aid of word boundaries during decoding.

We estimate the \textit{attention focus} (AF) of our model, that is, the likely time-position from which the model decodes a given output token.
We observe that attention patterns increase monotonically as tokens are decoded, so we heuristically define AF as the average attention weight position of all decoder layers and attention heads. 
When the AF shifts beyond a fixed proportion of our current attention window, we advance the window forward and proportionally truncate the decoder's context history.
This operation is repeated until we have decoded the entire episode.
A naive implementation of our algorithm often enters repetitive loops when encountering unintelligible speech or lyrical music not detected by our VAD, a known issue in attention models and neural text generation \cite{DBLP:journals/corr/abs-2002-05150, DBLP:conf/iclr/WelleckKRDCW20}.
We discover two patterns that arise when this happens: the number of n-gram repeats increases and the AF stops increasing.
We use an n-gram repetition detector and track AF changes to recover from these errors by pruning out repeating n-grams.
We use greedy search for unaligned evaluation due to memory limitations.


\subsection{Data Augmentation}
One issue with training speech models on well-aligned single utterances is that they cannot learn inter-utterance dependencies and adapt to imprecise utterance bounds.
We propose \textbf{ShiftAug}, which augments the dataset with random 10 to 30 second audio segments from episodes and trains the model to output the text of all utterances whose bounds lie within the sampled span.
We truncate the text of utterances that lie partially within these bounds proportional to the amount of intersecting audio.
This method is noisy because a sampled audio segment may not contain all the words in the utterance.
To address this issue, we introduce \textbf{AlignAug}, which uses heuristic forced-word alignments using the \texttt{Aeneas} tool\footnote{https://www.readbeyond.it/aeneas/}) to guide truncation.

\section{Results}

\subsection{Framework Comparison}
We report results from all models in \Cref{tab:experiments}.
We first compare the separate and joint frameworks for TAL aligned.
We find that separately trained ASR and SD models (Separate), when reconciled, obtain modest performance.
The same model trained jointly for ASR and SD (Joint) has a minor degradation in ASR performance---similar to findings in \cite{DBLP:conf/interspeech/ShafeySS19}---but the SD produced using speaker tokens is significantly worse than clustering embeddings from a separate SD model.
As our decoder learns embeddings for each vocabulary token, including each speaker token, we also tried clustering representations consisting of a weighted sum of speaker vocabulary embeddings for each utterance.
This method, however, yielded a few percent worse MWDE than using predicted speaker IDs directly.
We also trained an alternative joint setup where we used a separate speaker head in the decoder to classify speakers and treated speaker classification and ASR as a multitask loss, but this model was unable to converge.

We find that using clustering from an external diarization model confers significant benefits over merely using a joint model’s speaker predictions in the aligned case (SD+ column).
However, for the unaligned case, the joint model wins at diarization when the training is augmented by alignment training in the form of shifted augmentations.
Using our joint framework to determine utterance bounds reduces the unaligned MWDE to 62.2\%, a 29.1\% absolute reduction. 
Our results suggest that ASR and speaker identification may be conflicting tasks better suited for separately trained models, or may require more sophisticated multi-task learning schemes.

The presence of casing and punctuation in TAL contributes to its difficulty---we simulated the evaluation methodology of other ASR datasets (unifying casing and stripping punctuation) and observed 13.9\% and 38.2\% WER for the aligned and unaligned settings of our ShiftAug model, respectively.

\subsection{Pre-training}
We find that pre-training the acoustic model on LibriSpeech provides a 3.5\% absolute improvement to ASR performance in the aligned task, and a more pronounced 8.9\% improvement to ASR in unaligned.
This suggests that pre-training acoustic models on large audio corpora helps in learning useful features.
Our augmentation models builds upon this pre-training method.

\subsection{Unaligned Performance and Data Augmentation}
Overall, unaligned performance lags behind aligned performance by a significant margin even with pre-training.
In the unaligned setting, the separate framework relies on clustering speaker embeddings from the SD model (trained on speaker classification) for speaker change detection \cite{DBLP:conf/interspeech/YinBB18}.
This method performs poorly in diarization, with the separate framework achieving 91\% MWDE on unaligned.
Empirically, we find the SD model is unable to determine relative speaker boundaries on TAL unaligned, primarily due to highly variable microphone quality, lyrical music, and intra-conversation speaker diversity.
We find that using a jointly trained model to determine bounds, then averaging speaker embeddings with the utterances (Joint SD+) leads to much more stable predictions and more reasonable MWDE.
We hypothesize that this is because speaker embeddings in TAL at fine-granularity are noisy and order-less, making it difficult to cluster properly \cite{DBLP:conf/icassp/ZhangWZP019}.

A qualitative inspection suggests that accumulated VAD and utterance pruning errors contribute to the the disparity between aligned and unaligned performance.
Without data augmentation, our models perform poorly on TAL unaligned, as they are unable to learn inter-speaker utterance boundaries.
We find that ShiftAug is able to close some of the performance gap between unaligned and aligned results especially in diarization, and generally performs better in ASR than AlignAug likely due to the regularization from its noise.
Manual inspection of heuristic word boundaries from AlignAug reveals that the heuristic is overly conservative, at many times pruning excess tokens.
Augmentation improves joint prediction of speakers (SD), but we see no corresponding benefit when boosting with a separate SD model (SD+), suggesting that it improves speaker identification more than speaker boundary determination.

\begin{table}[]
\centering
\caption{Model performance on TAL test set. ASR and SD/SD+ are evaluated via WER and MWDE in percentages respectively. SD is computed directly from the joint model's speaker tokens, whereas SD+ uses clustering from external diarization model.}
\label{tab:experiments}
\begin{tabular}{@{}lllllll@{}}
\toprule
                      & \multicolumn{3}{c}{\textbf{Aligned}} & \multicolumn{3}{c}{\textbf{Unaligned}}               \\
 &
  \multicolumn{1}{l}{\textbf{ASR}} &
  \multicolumn{1}{l}{\textbf{SD}} &
  \multicolumn{1}{l}{\textbf{SD+}} &
  \multicolumn{1}{l}{\textbf{ASR}} &
  \multicolumn{1}{l}{\textbf{SD}} &
  \multicolumn{1}{l}{\textbf{SD+}} \\ \midrule

Separate                   & 24.3 & \multicolumn{1}{c}{--} & \textbf{15.4}  & 58.3 & \multicolumn{1}{c}{--} & 91.3 \\

\midrule

Joint     & 25.4  & 31.9 & 15.7 & 58.2 & 62.2 & \textbf{54.0} \\

+ Pre-training         & 21.9 & 29.5 & 15.7 & 49.3  & 63.8 & 54.6 \\

+ ShiftAug             & \textbf{18.9} & 29.1 & 15.6 & \textbf{42.1} & 38.2 & 55.8 \\

+ AlignAug           & 19.1 & \textbf{28.5} & 15.7 & 51.0 & \textbf{37.4} & 55.2   \\
\bottomrule
\end{tabular}
\vspace{-0.2in}
\end{table}

\section{Related Work}
\textbf{Joint ASR and SD:}
Joint ASR and SD within a single model was first explored by \cite{DBLP:conf/icassp/SarkarDNB18}, who spliced together audiobook snippets as a proxy for conversation.
\cite{DBLP:conf/asru/KandaHFXNW19} presented an alternating optimization strategy to jointly extract speaker embeddings and text in target-speaker transcription.
\cite{DBLP:conf/interspeech/ShafeySS19} expanded the work into two-party conversations but did not release their dataset.
Our work represents the first effort to jointly transcribe and diarize audio in a real-world setting with multiple speakers, punctuation, and casing, on a publicly available dataset.

\textbf{Pre-training:} Various approaches have been proposed for unsupervised pre-training of acoustic models to leverage large corpora of unlabeled audio, such as contrastive predictive coding \cite{DBLP:journals/corr/abs-1904-05862,DBLP:journals/corr/abs-1807-03748}, pseudo-labeling \cite{DBLP:journals/corr/abs-1911-08460}, and masked audio modeling \cite{DBLP:journals/corr/abs-1910-12638}.
We conduct supervised pre-training using labeled examples from Librispeech as we were unable to scale some of these more complex approaches to the TDS architecture in our preliminary experiments.

\textbf{Monotonic Attention:}
Traditional systems for transcribing long conversations rely on carefully-engineered pipelines and audio segmentation \cite{DBLP:conf/icassp/HainWBKDVGL07}.
Our method is more similar  to attention mechanisms that monotonically advance in time \cite{DBLP:journals/corr/abs-1712-05382,DBLP:journals/corr/abs-1909-12406,DBLP:journals/corr/abs-1811-05247}.
Our decoding algorithm is inspired by \cite{DBLP:conf/interspeech/MerboldtZSN19}, who show that the learned attention peak position is a good heuristic for advancing monotonic attention.

\section{Conclusion}
We present a new benchmark for ASR and SD in an extended multi-speaker conversational setting and explore three frameworks for learning both tasks.
We find that models that jointly learn ASR and SD perform best in the absence of known utterance bounds.
When bounds are provided, boosting with an external SD model improves diarization.
We introduce an algorithm that enables scaling ASR and SD to hour-long conversations and show significant performance improvement by incorporating pre-training and data augmentation methods.
TAL unaligned presents a new challenge for rich transcription of extended conversation.
We see opportunities for future work to investigate better pre-training and decoding algorithms.

\textbf{Acknowledgements:} This work was supported in part by NSF awards CNS-1730158, ACI-1540112,  ACI-1541349, OAC-1826967, IIS-1750063, the UC Office of the President, the UCSD's California Institute for Telecommunications and Information Technology/Qualcomm Institute, CENIC for the 100Gpbs networks and the Alexa Prize Grand Challenge 3.

\bibliographystyle{IEEEtran}

\bibliography{bib}

\end{document}